\documentclass[aps,prl,nofootinbib,longbibliography,superscriptaddress,twocolumn]{revtex4-2}
\usepackage{amsmath}
\usepackage{amssymb}
\usepackage{graphicx}
\usepackage{dcolumn}
\usepackage{bm}
\usepackage[colorlinks=true,linkcolor=blue,anchorcolor=blue, citecolor=blue,urlcolor=blue]{hyperref}
\usepackage[mathlines]{lineno}
\usepackage{ulem}
\usepackage{siunitx}
\usepackage{gensymb}

\makeatother

\begin{document}
	
\title{Spontaneous Fully Compensated Ferrimagnetism}
\author{Bingbing Wang}
\affiliation{Centre for Quantum Physics, Key Laboratory of Advanced Optoelectronic Quantum Architecture and Measurement (MOE), School of Physics, Beijing Institute of Technology, Beijing 100081, China}
\affiliation{Beijing Key Laboratory of Quantum Matter State Control and Ultra-Precision Measurement Technology, Beijing 100081, China}
	
\author{Yongpan Li}
\affiliation{Centre for Quantum Physics, Key Laboratory of Advanced Optoelectronic Quantum Architecture and Measurement (MOE), School of Physics, Beijing Institute of Technology, Beijing 100081, China}
\affiliation{Beijing Key Laboratory of Quantum Matter State Control and Ultra-Precision Measurement Technology, Beijing 100081, China}

\author{Yichen Liu}
\affiliation{Centre for Quantum Physics, Key Laboratory of Advanced Optoelectronic Quantum Architecture and Measurement (MOE), School of Physics, Beijing Institute of Technology, Beijing 100081, China}
\affiliation{Beijing Key Laboratory of Quantum Matter State Control and Ultra-Precision Measurement Technology, Beijing 100081, China}
	
\author{Cheng-Cheng Liu}
\email{ccliu@bit.edu.cn}
\affiliation{Centre for Quantum Physics, Key Laboratory of Advanced Optoelectronic Quantum Architecture and Measurement (MOE), School of Physics, Beijing Institute of Technology, Beijing 100081, China}
\affiliation{Beijing Key Laboratory of Quantum Matter State Control and Ultra-Precision Measurement Technology, Beijing 100081, China}

\begin{abstract}
We propose a general mechanism for the spontaneous emergence of filling-enforced fully compensated ferrimagnetism (fFIM), characterized by zero net magnetization yet ferromagnetic-like spin-split band structures. Using Hartree-Fock mean-field calculations of the Hubbard model, we map out the stability regime of spontaneous fFIM over a broad parameter space of interaction strength and staggered potential. We show the unique quantum-geometry-governed optical selection rules and the abundant valley- and spin-related physics of electronics and optics arising from the emergence of fFIM order, with tunable spin-polarized and valley-contrasting charge and spin currents. Furthermore, based on our theory, we demonstrate that spontaneous fFIM can be realized in nominally nonmagnetic graphene via defect engineering. Our results establish a unified framework for the mechanism, emergent properties, and materials realization of spontaneous fFIM, opening new opportunities for spintronic, valleytronic, and optoelectronic applications.
\end{abstract}

\maketitle
\textit{Introduction.}---
Collinear fully compensated magnetism can generally be classified into two categories~\cite{PhysRevLett.74.1171, PhysRevLett.97.026401,wurmehl2006valence,PhysRevLett.134.116703,ZhangPRX2025,RevModPhys.90.015005,PhysRevX.12.040002,PhysRevX.12.040501}. The first includes conventional antiferromagnetism (AFM) and recently identified altermagnetism (AM)~\cite{RevModPhys.90.015005,PhysRevX.12.040002,PhysRevX.12.040501}, which belong to \textit{symmetry-enforced} compensated magnets, where crystal symmetries relate opposing magnetic sublattices. The second category is \textit{filling-enforced} fully compensated ferrimagnetism (fFIM)~\cite{PhysRevLett.134.116703,PhysRevLett.74.1171}, characterized by the absence of symmetry operations connecting magnetic sublattices together with an appropriate electron filling that guarantees zero net magnetization. This lack of symmetry constraint allows spin splitting in the electronic band structure despite complete magnetic compensation.
The filling-enforced fully compensated ferrimagnets (fFIMs) combine key characteristics of both ferromagnets (FMs) and antiferromagnets (AFMs), offering an appealing platform for spintronics~\cite{nayak2015design,siddiqui2018current,liensberger2019exchange,zhou2021efficient,kim2022ferrimagnetic,wang2024electrical}.  Similar to FMs, fFIMs can support fully spin-polarized electrical currents~\cite{PhysRevLett.74.1171}, while their compensated magnetic order inherits the hallmark advantages of AFMs, including ultrafast spin dynamics, robustness against external magnetic fields, and vanishing stray fields~\cite{RevModPhys.90.015005}. Moreover, the breaking of combined space-time inversion ($\mathcal{PT}$) symmetry in fFIMs generates finite Berry curvature, enabling anomalous transport phenomena such as anomalous Hall, Nernst, and magneto-optical effects~\cite{PhysRevB.111.195154,guoSlidingFerroelectricMetal2025,ChenUnconventionalMagnetism,liTunableTopologicalSuperconductivity2025}.

A number of candidate materials and engineering strategies for realizing fFIM have recently been proposed~\cite{PhysRevB.57.10613,nie2008possible,vzic2016designing,semboshiNewTypeHalfmetallic2022,PhysRevLett.132.156502,9syc-71w8,zhang2025electricfield}, particularly in layered materials and heterostructures~\cite{PhysRevLett.134.116703,yu2025electrically,Guo2025AchievingFC,6wmg-hxdq}. 
However, nearly all predicted realizations rely on magnetic ions (e.g., Cr or Mn), where magnetism originates from localized $d$- or $f$-electron moments. In contrast, the spontaneous emergence of fFIM driven purely by electronic correlations in initially nonmagnetic systems remains largely unexplored. Recent observations of interaction-induced magnetism in $p$-electron systems~\cite{songJanusGrapheneNanoribbons2025,PhysRevLett.132.046201,6tny-vt8q} suggest a promising alternative route. Understanding the general mechanism of spontaneous fFIM, uncovering its emergent novel physical properties, and establishing feasible realization strategies therefore constitute an important and timely challenge.


In this work, we propose a general mechanism and material realization strategy for interaction-driven spontaneous fFIM.
Using Hartree-Fock mean-field calculations of the Hubbard model on bipartite lattices, including honeycomb and square geometries, we map out the phase diagrams as functions of interaction strength and staggered potential and identify a robust spontaneous fFIM phase. We show the novel optical selection rules based on quantum geometry and the abundant valley- and spin-related physics in electronics and optics arising from the fFIM order, with tunable spin-polarized and valley-contrasting charge and spin currents. Finally, we propose a feasible route to realizing spontaneous fFIM in graphene-based systems via defect engineering. By establishing both a general mechanism and principles of design and control, our work advances a new paradigm for compensated magnetism beyond magnetic-ion systems, opening access to previously unexplored material platforms and functionalities.





\textit{Spontaneous fFIM model.}---We propose a general model for the spontaneous fFIM by considering the bipartite lattice with the onsite Hubbard interaction. The model Hamiltonian can be written in the form

\begin{equation}\label{eq:TotHam}
	H=-\sum_{\langle i,j\rangle,\sigma}t_{ij}(\hat{c}_{i,\sigma}^{\dag}\hat{c}_{j,\sigma}+h.c.)+\sum_{i,\sigma}u_{i}\hat{n}_{i,\sigma}+\sum_{i}U\hat{n}_{i\uparrow}\hat{n}_{i\downarrow}.
\end{equation}
Here, the first term is the nearest electron hopping. $u_i$ is the on-site potential, and the last term represents the local Hubbard interaction.  $\hat{n}_{i\sigma}=\hat{c}_{i,\sigma}^{\dag}\hat{c}_{j,\sigma}$ is the density operator. After Fourier's transformation and mean field decoupling, with the details in Supplemental Materials (SM)~\cite{supplemental}, the mean field Hamiltonian at each $\boldsymbol{k}$ point in Brillouin zone (BZ)  is expressed as

\begin{equation}\label{eq:TotmreanHam}
	\begin{aligned}
	H_{\mathrm{MF}}(\boldsymbol{k})=&T_{1}(\boldsymbol{k})\sigma_{x}+T_{2}(\boldsymbol{k})\sigma_{y}+\frac{1}{2}\Delta\sigma_{z}\\
	&-U\delta m s_{z}\sigma_{z}+const,
	\end{aligned}
\end{equation}
where $s$ and $\sigma$ denote the spin and sublattice degrees of freedom, respectively. To confirm the formation of staggered magnetization, we introduce the order parameter $\delta m=(-1)^{\alpha_i}\frac{\langle n_{i\uparrow}\rangle-\langle n_{i\downarrow}\rangle}{2}$, where $\alpha_{A}=0$ and $\alpha_{B}=1$. The total electron density is denoted as $n=\sum_{s}\langle n_{As}+n_{Bs}\rangle$. The electron density on each sublattice can be expressed as $\langle n_{As}\rangle=\frac{n}{2}+s\delta m$, $\langle n_{Bs}\rangle=\frac{n}{2}-s\delta m$, where $s=1$ for spin-up and -1 for spin-down. The electron density for each spin channel reads $n_{s}=\langle n_{As}\rangle+\langle n_{Bs}\rangle$. We discuss two typical types of 2D bipartite lattices---the honeycomb lattice and the square lattice.
For the honeycomb case $T_{1}=t[1+2\mathrm{cos}(\frac{\sqrt{3}}{2}k_{y})\mathrm{cos}(\frac{1}{2}k_{x})]$ and $T_{2}=2t\mathrm{sin}(\frac{\sqrt{3}}{2}k_{y})\mathrm{cos}(\frac{1}{2}k_{x})$. For the square lattice, $T_{1}=2t(\mathrm{cos}(\frac{\sqrt{2}}{2}k_{x})+\mathrm{cos}(\frac{\sqrt{2}}{2}k_{y}))$ and $T_{2}=0$. Here, $\Delta=u_{A}-u_{B}$ represents the magnitude of the staggered potential. We set $u_B=0$ thus $\Delta=u_A$. The final constant term, given by $\frac{\Delta}{2}+U(\frac{n}{2}-\frac{n^{2}}{4}+\delta m^{2})$, leads to overall shift in energy. The eigenvalues of  $H_{\mathrm{MF}}(\boldsymbol{k})$ can be expressed analytically as $\mathrm{E}_{s}=\pm\sqrt{\epsilon_{\boldsymbol{k}}^{2}+(\frac{\Delta}{2}-s U\delta m)^{2}}$, where $\epsilon_{\boldsymbol{k}}^{2}=|T_{1}(\boldsymbol{k})|^{2}+|T_{2}(\boldsymbol{k})|^{2}$.
\begin{figure}[t]
	\includegraphics[width=0.5\textwidth]{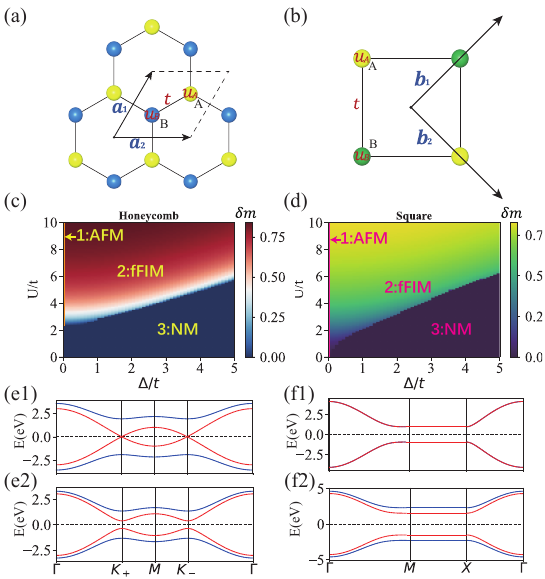}
	\caption{Spontaneous fully compensated ferrimagnetism in Hubbard models. (a) Honeycomb lattice model. The honeycomb lattice vectors are defined as $\boldsymbol{a_{1}}=(1/2,\sqrt{3}/2)$ and $\boldsymbol{a_{2}}=(1,0)$. (b) Square lattice model. The square lattice model are defined as $\boldsymbol{b_{1}}=1/\sqrt{2}(1,1)$ and $\boldsymbol{b_{2}}=1/\sqrt{2}(1,-1)$. The onsite energies $u_A$ and $u_B$ are applied on sublattices A and B. (c)-(d) The Hartree-Fock mean-field phase diagrams in the $U/t$-$\Delta/t$ plane for (c) honeycomb lattice and (d) square lattice at half-filling. AFM, fFIM, NM denote conventional antiferromagnetism, filling-enforced fully compensated ferrigmanetism and non-magnetism, respectively. (e1)-(e2) Band spectra of half-metallic and gapped fFIM honeycomb model. The parameters are chosen to be $U=5t$, $\Delta=3.89t$ in (e1) and $U=5t$, $\Delta=4.1t$ in (e2). (f1)-(f2) Band spectra of AFM and fFIM square lattice. The parameters are chosen to be $U=3.2t$, $\Delta=0$ in (f1) and $U=5t$, $\Delta=t$ in (f2). The red and blue colors represent the spin-up and spin-down components, respectively.}\label{Modelphase}
\end{figure}

We focus on the half-filling case with $2n=2$. Through self-consistent Hartree-Fock iterations, we obtain the magnetic phases as functions of the Hubbard interaction $U$ and staggered potential $\Delta$, as shown in the phase diagram Figs.~\ref{Modelphase}(c)(d). For the honeycomb lattice at $\Delta=0$, the density of states vanishes at the Fermi energy. As a result, a critical value $U_c$ is needed for the ordered phase. When $U$ exceeds $U_c\approx2.23t$, which is consistent with previous studies~\cite{PhysRevB.101.125103,sorellaSemiMetalInsulatorTransitionHubbard1992}, both $\mathcal{P}$ and $\mathcal{T}$ symmetries are spontaneously broken, leading to staggered spin polarization on each sublattice. Nonetheless, the combined symmetry $\mathcal{PT}$ remains preserved, thereby stabilizing the conventional AFM state, as indicated by the vertical orange line in Fig.~\ref{Modelphase}(c). In contrast, the square lattice exhibits a divergent density of states near the Fermi level and possesses a perfectly nested Fermi surface satisfying $\epsilon_{\boldsymbol{k}}=\epsilon_{\boldsymbol{k}+(\pi,\pi)}$~\cite{PhysRevB.31.4403,PhysRevB.40.506}. Consequently, an infinitesimally small $U$ is sufficient to yield a solution with $\delta m\neq 0$, as demonstrated in Fig.~\ref{Modelphase}(d). As an illustration of the resulting ordered phase, we plot the band structure of the square lattice in the AFM state in Fig.~\ref{Modelphase}(f1).

\begin{figure}[t]
	\includegraphics[width=0.5\textwidth]{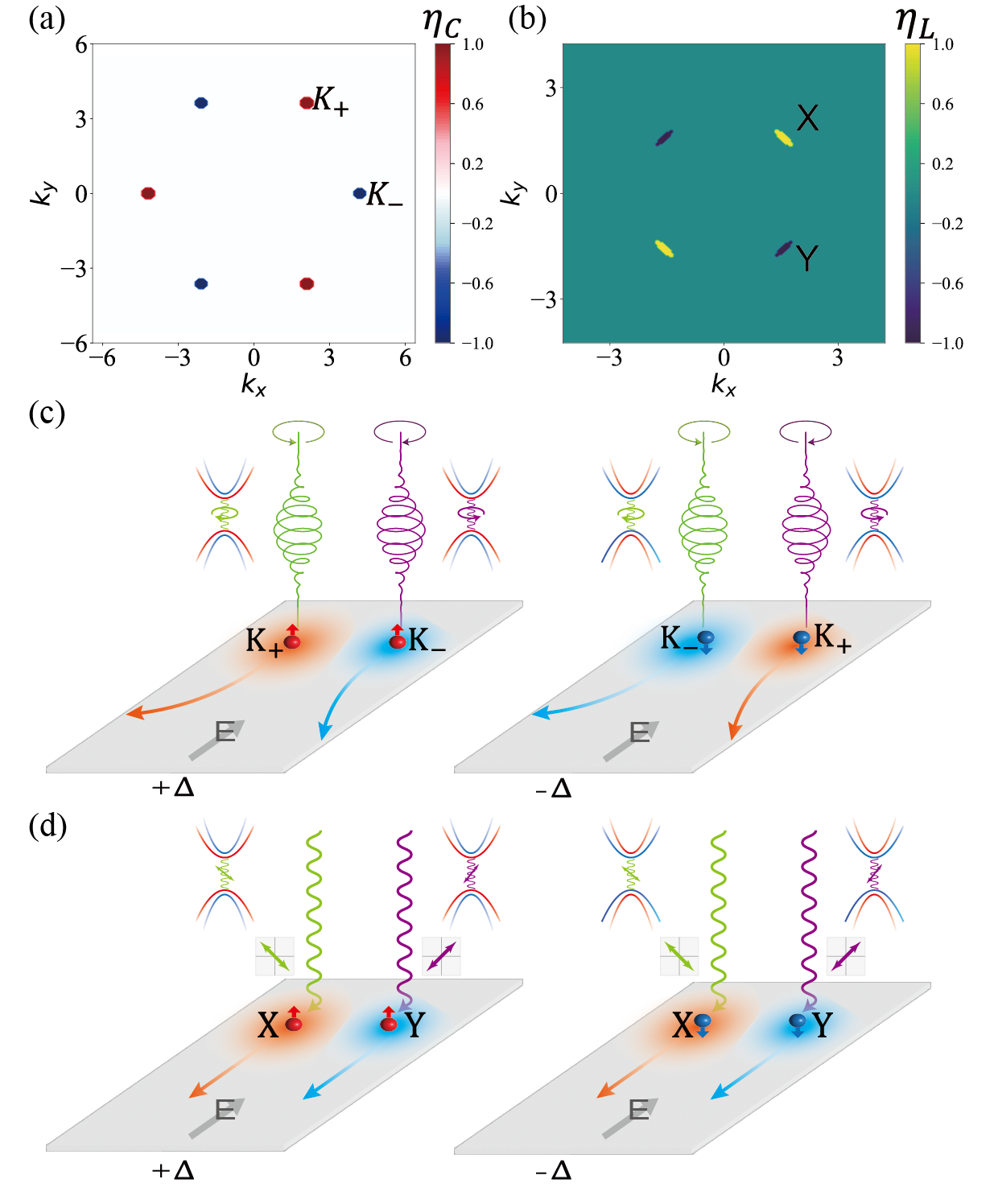}
	\caption{Circular and linear polarized optical selection rules in spontaneous fFIMs. (a) Degrees of circular polarization in the honeycomb lattice. (b) Degrees of linear polarization in the square lattice. In this case, additional next nearest neighbor hopping is included, which induces two valleys in the energy spectrum. (c) Valley-contrasted selection rules of circularly polarized light in $K_{+}$ and $K_{-}$ valleys. The Berry curvatures of the two valleys are opposite in sign, leading to opposite circular polarization selectivity. Excited electrons exhibit the same longitudinal and opposite transverse currents. Reversing the staggered potential switches spin polarization and the valley selection for a specific circular polarization of light. (d) Valley-contrasted selection rules of linearly polarized light. X and Y denote two valleys with opposite quantum metric. The Berry curvatures are zero, with zero transverse current. The spin-up and spin-down in the band structure are labeled in red and blue. Holes alongside the excited electrons, possessing both opposite velocity and opposite spin relative to their electronic counterparts. The parameters are taken as $U=4t$ and $\Delta=2t$ for (a) and $U=4.8t$ and $\Delta=0.7t$ for (b).}\label{Light select}
\end{figure}

When $\Delta \neq 0$ and under strong U, the system develops staggered magnetic moments and loses the combined $\mathcal{PT}$ symmetry simultaneously. As a result, no symmetry that connects the magnetic moments of sublattices A and B remains. In this case, the spin-up and spin-down bands split throughout the whole BZ, as shown in Figs.~\ref{Modelphase}(e1)-(e2) and (f2), while yielding a zero net magnetic moment with $n_{\uparrow}=n_{\downarrow}$. From the eigenenergy expression, it can be seen that when the staggered potential and interaction satisfy the condition $\Delta=\pm2U\delta m$, the system reaches a half-metallic state. In this state, the band for one spin channel closes, whereas the band for the opposite spin channel remains gapped with a magnitude of $2\Delta$, which is verified in Fig.~\ref{Modelphase}(e1). Such half-metallic fFIMs can generate 100\%  spin-polarized current. The filling-enforced fFIM occurs when electron filling satisfies $n_{\uparrow}=n_{\downarrow}$, ensuring zero net magnetization while maintaining a gapped spin channel. This compensation is robust under perturbations as long as the gap remains open, independent of symmetry constraints or external fields, due to the integer quantization of electron numbers in each spin channel. Such gap-guaranteed quantization ensures full compensation with a robust zero magnetization regardless of whether the system is an insulator or a half-metal.

Observing the phase diagrams in Figs.~\ref{Modelphase}(c)(d), we note a competitive relationship between the interaction $U$ and the on-site energy $\Delta$. For a given $U$, gradually increasing $\Delta$ drives a phase transition from the AFM to the fFIM. At sufficiently large $\Delta$, the ground state of the system becomes a non-magnetism (NM) phase. This can be understood through a relatively straightforward physical picture: when the energy at sublattice A becomes too high, electrons preferentially occupy the lower-energy sublattice B. In this scenario, sublattice A hosts no electrons, while the spin-up and spin-down electrons on sublattice B cancel each other out.

\textit{Optical selection rules in spontaneous fFIMs.}---Using optical selection rules as examples, we show the abundant valley- and spin-related physics in optics, and electronics arising from the spontaneous emergence of fFIM order. The optical selection rules for linearly and circularly polarized light are characterized by their respective degrees of polarization~\cite{bdhy-hnd2,PhysRevB.77.235406}, i.e.,
\begin{align}
	\eta_{L}(\boldsymbol{k})&=\frac{2g_{vc}^{xy}(\boldsymbol{k})}{g_{vc}^{xx}(\boldsymbol{k})+g_{vc}^{yy}(\boldsymbol{k})},\label{eq:select_rule_lin}\\
	\eta_{C}(\boldsymbol{k})&=-\frac{\Omega_{vc}^{xy}(\boldsymbol{k})}{g_{vc}^{xx}(\boldsymbol{k})+g_{vc}^{yy}(\boldsymbol{k})},\label{eq:select_rule_cir}
\end{align}
with $g_{vc}^{\alpha\beta}(\boldsymbol{k})=\mathrm{Re}(v_{vc,\boldsymbol{k}}^{\alpha}v_{cv,\boldsymbol{k}}^{\beta})/{\omega_{vc,\boldsymbol{k}}^{2}}$ and $\Omega_{vc}^{\alpha\beta}(\boldsymbol{k})=2\mathrm{Im}(v_{vc,\boldsymbol{k}}^{\alpha}v_{cv,\boldsymbol{k}}^{\beta})/{\omega_{vc,\boldsymbol{k}}^{2}}$. The quantum metric $g_{vc}^{\alpha\beta}(\boldsymbol{k})$ and Berry curvature $\Omega_{vc}^{\alpha\beta}(\boldsymbol{k})$ are the real and imaginary parts of the quantum geometric tensor, respectively. The subscript $v\ (c)$ denotes the valence (conduction) band. $v_{vc,\boldsymbol{k}}^{\alpha}$ is the matrix element of the velocity operator in the $\alpha$ direction, and $\hbar\omega_{vc,\boldsymbol{k}}$ is the energy difference between the $v$-th and $c$-th band at the $\boldsymbol{k}$ point, which corresponds to the optical excitation energy.

The two valleys in the honeycomb lattice model of spontaneous fFIM possess three-fold rotational symmetry and are connected by the spin group symmetry operator $[C_2||\mathcal{T}]$ (the time-reversal symmetry combined with the spin flipping). Since three-fold rotational symmetry forbids the finite quantum metric~\cite{bdhy-hnd2} and the Berry curvature is odd under $[C_2||\mathcal{T}]$, the two valleys exhibit valley-contrasting optical selection rules for circular polarization in the optical transition at the valleys. As shown in Fig.~\ref{Light select}(a), the numerical calculations indicate that the $K_+$ and $K_-$ valleys have opposite degrees of circular polarization. Due to the presence of $[C_2||\mathcal{T}]$, the $K_+$ and $K_-$ valleys in fFIMs are locked with the same spin. This mechanism is fundamentally different from the case of spin-valley locking in transition metal dichalcogenides, in which time-reversal symmetry $\mathcal{T}$ requires that the spin splitting at the two different valleys must be opposite~\cite{PhysRevLett.108.196802}, and consequently leads to entirely different optical selection rules. Moreover, the type of spin that is associated with the valleys can be tuned by the stagger potential. For the positive (negative) stagger potential, the two valleys are locked with up (down) spin.

The unique and highly tunable nature of the valley-contrasted optical selection rules in the honeycomb lattice model paves the way for novel spintronic and valleytronic applications in fFIMs. Under left (right) circularly polarized light, electrons and holes are excited in the $K_{+}$ ($K_{-}$) valley with a specific spin polarization. Applying in an in-plane electric field induces a transverse velocity on these electrons that is opposite for the two valleys, as illustrated in Fig.~\ref{Light select}(c), in which the left and right panels under opposite staggered potential are connected by $\mathcal{PT}$ symmetry. Within the same valley, the holes move with a velocity opposite to that of the electrons. Remarkably, reversing the staggered potential, for example, by applying an external electric field, inverts both the valley and spin polarization of the optically excited carriers for a specific circular polarization, while leaving the deflection direction of the carrier unchanged. Therefore, the chirality of circular light governs the carrier transverse direction, whereas the staggered potential dictates the spin and valley polarization.

For the square lattice, we introduce next-nearest neighbour (NNN) hoppings (See SM~\cite{supplemental} for details), generating two valleys at the X and Y points, where the quantum metric assumes non-zero values. These valleys are connected by mirror ($\mathcal{M}$) symmetry, and thus possess opposite quantum metric and opposite degree of linear polarization. As illustrated in Fig. \ref{Light select}(b), the numerical calculations reveal that the opposite values of $\eta_{L}$ emerge in these two valleys, and the result is in agreement with the symmetry analysis. Therefore, optical transitions at the two valleys have different selection rules: light with orthogonal linear polarizations selectively excites carriers in the corresponding valley only, as shown in Fig.~\ref{Light select}(d). Moreover, applying the opposite staggered potential can reverse the spin of the carriers. Since the two valleys lie on $\mathcal{M}$-invariant lines, the Berry curvature and transverse current vanish. Reversing $\Delta$ is approximately equivalent to applying a $\mathcal{PT}$ operation to the model. The longitudinal current comprises the charge current $J_c$ and spin current $J_s$. Under the $\mathcal{PT}$ operation, $J_c$ remains unchanged, while the sign of $J_s$ is reversed. Therefore, reversing staggered potential flips the spin polarization of carriers and spin current $J_s$ under the same linearly polarized optical excitation, as illustrated in Fig.~\ref{Light select}(d).

From the low-energy $k\cdot p$ model, we can analytically analyze the degree of polarization. Expanding the Hamiltonian at the $K_{+}/K_{-}$ valley for honeycomb lattice yields $H_{K_{+}/K_{-}}=-\hbar v_{F}(k_{x}\sigma_{x}+\xi k_{y}\sigma_{y})+(\frac{1}{2}\Delta\mp U\delta m)\sigma_{z}$ with $v_{F}=\frac{3t}{2\hbar}$, and X/Y valley for square lattice yields $H_{X/Y}=-t(k_{x}+\xi k_{y})\sigma_{x}+(\frac{1}{2}\Delta\mp U\delta m)\sigma_{z}$, where $\xi=1$ for the $K_{+}$ (X) valley and $-1$ for the $K_{-}$ (Y) valley. For honeycomb lattice, the matrix representations of velocity operators $M_{v^{x}}=-v_{F}\sigma_{x}$ and $M_{v^{y}}=-v_{F}\xi \sigma_{y}\mathrm{sign}(\Delta)$, while for square lattice $M_{v^{x}}=-t\sigma_{x}$ and $M_{v^{y}}=-\xi t\sigma_{x}$. From Eqs. (\ref{eq:select_rule_lin}) and (\ref{eq:select_rule_cir}), we can derive that $\eta_{C}(K_{-})=-\eta_{C}(K_{+})=\mathrm{sign}(\Delta)$, and $\eta_{L}(X)=-\eta_{L}(Y)=1$, with a detailed derivation provided in the SM~\cite{supplemental}. This observation can be understood from the distinct responses of $\Omega$ and $g$ to the $\mathcal{PT}$ operation: circular polarization, which depends on $\Omega$, reverses sign under $\mathcal{PT}$, whereas linear polarization, which depends on $g$, does not. These findings are in good agreement with our numerical results and schematic depiction in Fig.~\ref{Light select}.

\textit{Candidate materials with non-magnetic elements.}---There have been reports on the emergence of magnetism in $\pi$-bonded $p$-orbital electrons, such as in graphene~\cite{tang2020magnetic,PhysRevB.75.125408,PhysRevB.77.195428,DirectImagingMeyer,PhysRevLett.104.096804,BuildingUnconventionalBausa,gonzalez-herreroAtomicscaleControlGraphene2016,PhysRevLett.132.046201} and nitrogen-containing materials~\cite{PhysRevLett.108.197207}. Compared to magnetic atoms with $d$ or $f$ electrons, C/N atoms are lighter, exhibiting weaker spin-orbit coupling and long spin coherence time, thereby allowing more effective tuning. This naturally raises the question: Can spontaneous fFIM emerge in these non-magnetic materials?

\begin{figure}[t]
	\includegraphics[width=0.5\textwidth]{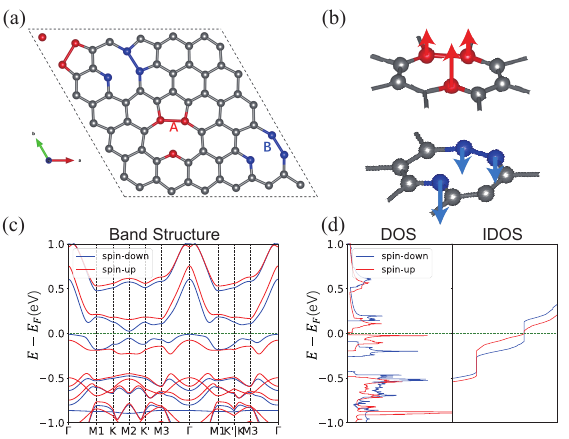}
	\caption{Spontaneous fFIM in graphene with vacancies. (a) The 6$\times$6 graphene supercell with two pairs of vacancies. The red regions represent spin-up components, and the blue regions represent spin-down components. (b) Illustration of magnetic distribution. Magnetic moments are predominantly localized on the carbon atoms containing an unpaired $\pi$-electron. (c) Band structure. (d) Density of states (DOS) and integrated density of states (IDOS). From the DOS, we can see that the Fermi energy level lies in the gap. The IDOS diagram reveals that the spin-up and spin-down electronic states at the Fermi level exhibit full compensation.}\label{Material}
\end{figure}

Here, we present the strategy for designing and inducing fFIMs in material systems that are intrinsically non-magnetic. Based on the preceding model analysis, we propose that the material should ideally possess a bipartite lattice structure. Another prerequisite is the introduction of an effective staggered potential to break $\mathcal{P}$ symmetry. These naturally direct our focus to graphene, the quintessential bipartite two-dimensional material, in which magnetism control has long been a subject of scientific interest. Methods to induce magnetism in graphene include introducing defects~\cite{PhysRevB.75.125408,PhysRevB.77.195428,DirectImagingMeyer,PhysRevLett.104.096804}, or hydrogenating specific carbon atoms~\cite{BuildingUnconventionalBausa,gonzalez-herreroAtomicscaleControlGraphene2016,PhysRevLett.132.046201}. These approaches break the $\pi$-bonds in one sublattice (e.g., A), thereby isolating the $p_z$ orbitals of the carbon atoms on the other sublattice (B) and creating localized magnetic moments. In this work, we employ a strategy of introducing vacancies into graphene, which is experimentally viable~\cite{ZHANG2020194,PhysRevLett.104.096804,banhartStructuralDefectsGraphene2011,liuElementalSuperdopingGraphene2016}, with the specific requirement that the vacancies are created on both the A and B sublattices in equal number. According to the second Lieb's theorem~\cite{PhysRevLett.62.1201}, the total spin $S=\frac{1}{2}(n(A)-n(B))=0$, suggesting a potential antiferromagnetic ground state.  However, introducing just one A-B vacancy pair is insufficient to break $\mathcal{P}$, and the resulting system would remain the conventional $\mathcal{PT}$ antiferromagnetism. Therefore, a minimum of two such vacancy pairs is required. By arranging these vacancies in a configuration that simply breaks $\mathcal{P}$, as illustrated in Fig.~\ref{Material}(a), we are able to realize the fFIM state.

By employing DFT calculations (see the SM~\cite{supplemental} for computational details), we investigated 6$\times$6 supercell containing two vacancy pairs as an example. After atomic relaxation, the structure reconstructs such that around each vacancy, two of the three surrounding carbon atoms form a bond, creating a pentagon and leaving the remaining carbon atom with an unpaired electrons in its $p_z$ orbital. Our numerical results reveal that the local magnetic moments around the A-sublattice vacancies are opposite to those around the B-sublattice vacancies. The spatial distribution of these magnetic moments, plotted in Fig.~\ref{Material}(b), shows that the primary moments reside on the two-coordinated carbon atom, with weaker moments induced on other nearby atoms. The electronic band structure in Fig.~\ref{Material}(c) exhibits a ferromagnetic-like spin splitting. Furthermore, the integrated density of states (IDOS) in Fig.~\ref{Material}(d) demonstrates that the number of spin-up and spin-down states below the Fermi level is equal, indicating full compensation. Thus, by solely engineering the positions of defects in non-magnetic graphene, we have realized spontaneous fFIM. Moreover, by arranging the defects to preserve specific symmetries, such as $\mathcal{C}_{4z}\mathcal{T}$ or $\mathcal{MT}$, one can realize another unconventional magnetic phase---altermagnetism, showing that our scheme provides a powerful theoretical guide for fFIM design.

\textit{Discussion.}---We propose a general mechanism for the spontaneous emergence of fFIM with zero net magnetization. The global phase diagram with respect to $U$ and $\Delta$ is obtained within the mean-field framework. Depending on optical selection rules (both circular and linear), these fFIMs exhibit a highly tunable, valley-selective, spin-polarized longitudinal and transverse current excited by specific polarized light. This direct coupling among spin, valley, and optical polarization degrees of freedom underscores the potential of spontaneous fFIMs as a unified platform for multi-functional quantum devices bridging spintronics, valleytronics, and optoelectronics.

Our scheme provides a universal guideline for the spontaneous fFIM magnetic order, which can be experimentally realized in non-magnetic graphene with defects. This strategy is readily transferable to other monolayer Group-IV graphene-like materials (e.g., silicene, germanene, and stanene), and three-dimensional layered materials. We further propose that our scheme could be extended to other low-dimensional carbon-based organic systems. For instance, introducing aromatic benzene rings or clar sextets into super-zethrene structures~\cite{D0CC02513E}—or alternatively, creating benzene-ring defects on one side of graphene nanoribbons~\cite{On-SurfaceSynthesisTurco,songJanusGrapheneNanoribbons2025}—can break the symmetry between sublattices, thereby offering a feasible route toward realizing spontaneous fFIM. The interplay between orbital and spin degrees of freedom gives rise to spin order and orbital order, leading to altermagnetic behavior~\cite{PhysRevLett.132.236701,cjzw-j4v7}. Beyond spin magnetic moments, orbital magnetism also represents a promising avenue for future exploration in the pursuit of spontaneous fFIM systems.

\begin{acknowledgments}
	\textit{Acknowledgments.}--- We thank Yilin Wang for stimulating discussions. The work is supported by the Science Fund for Creative Research Groups of  NSFC (Grant No. 12321004), the NSF of China (Grant No.~12374055), and the National Key R\&D Program of China (Grant No.~2020YFA0308800).
\end{acknowledgments}

\bibliography{reference}
\end{document}